# The state of quantum computing applications in health and medicine

Frederik F. Flöther

IBM Quantum, IBM Research, Rüschlikon, Switzerland and QuantumBasel, uptownBasel Infinity Corp., Arlesheim, Switzerland

**Abstract**

Medicine, including fields in healthcare and life sciences, has seen a flurry of quantum-related activities and experiments in the last few years (although biology and quantum theory have arguably been entangled ever since Schrödinger's cat). The initial focus was on biochemical and computational biology problems; recently, however, clinical and medical quantum solutions have drawn increasing interest. The rapid emergence of quantum computing in health and medicine necessitates a mapping of the landscape. In this review, clinical and medical proof-of-concept quantum computing applications are outlined and put into perspective. These consist of over 40 experimental and theoretical studies. The use case areas span genomics, clinical research and discovery, diagnostics, and treatments and interventions. Quantum machine learning (QML) in particular has rapidly evolved and shown to be competitive with classical benchmarks in recent medical research. Near-term QML algorithms have been trained with diverse clinical and real-world data sets. This includes studies in generating new molecular entities as drug candidates, diagnosing based on medical image classification, predicting patient persistence, forecasting treatment effectiveness, and tailoring radiotherapy. The use cases and algorithms are summarized and an outlook on medicine in the quantum era, including technical and ethical challenges, is provided.

## Introduction

Quantum computing hardware and software have made enormous strides over the last years (Gill et al., 2022). Questions around quantum computing's impact on research and society have changed from "if" to "when/how". The 2020s have been described as the "quantum decade" (Sieger et al., 2023), and the first production solutions that drive scientific and business value are expected to become available over the next years. Thus, a cross-industry race has begun to secure quantum talent, build quantum skills, map real-world problems to quantum algorithms, capture quantum application intellectual property (IP), and prepare for quantum advantages. Certain types of applications gathered research interest right from the start; for instance, simulating nature through enhanced chemistry and physics calculations (Daley et al., 2022) and solving finance problems (Herman et al., 2022).

In healthcare and life sciences, the initial focus was on biochemical and computational biology problems (Emani et al., 2021; Outeiral et al., 2021; Fedorov and Gelfand, 2021; Marchetti et al., 2022; Cordier et al., 2022; Baiardi et al., 2022). Recently, the possibilities of quantum computing have increasingly sparked research interest in other fields as well. This is evidenced by clinical and medical proof-of-concept studies, which have seen a remarkable growth over the last years in conjunction with the exploration of use cases in healthcare (Flöther et al., 2022a), medicine (Maniscalco et al., 2022), and life sciences (Flöther et al., 2022b).

Defined by the characteristics of the algorithms and the types of problems for which the algorithms are used, three primary quantum algorithm application categories can generally be distinguished:

1. Simulating nature – including chemistry, materials science, and physics
2. Processing data with complex structure – including artificial intelligence / machine learning (AI/ML), factoring, and ranking
3. Search and optimization – including pricing, risk analysis, and sampling

Note that a given quantum algorithm may be part of more than one category. For example, the variational quantum eigensolver (VQE) algorithm (Tilly et al., 2022) has been applied to strongly correlated systems in chemistry ("Simulating nature") as well as finding the optimal configuration of nonquantum systems ("Search and optimization").

A common misconception about quantum computing is that the hardware and software are very similar to their classical counterparts. This is not the case. In fact, quantum algorithms leverage the principles of quantum mechanics, including quantum entanglement, interference, and superposition, in order to tackle problems in novel ways. For a classical computer, the computational power is closely related to the number of transistors. On the other hand, for a quantum computer the number of quantum basis states that can be explored and manipulated in a calculation doubles with the addition of each qubit, thus growing exponentially. This exponential growth underlies the power of quantum algorithms and enables a range of use case-dependent benefits, which may include one or more of:

- Accuracy (e.g., of an AI/ML model)
- Calculation speed
- Energy efficiency
- Input data requirements (quality, volume)

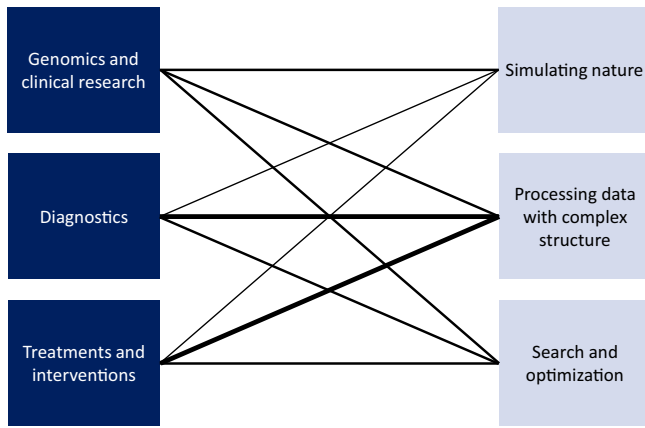

**Figure 1.** Three key quantum computing use case areas in health and medicine linked to quantum algorithm application categories. The wider the connecting line, the more applicable the category.

## Results

The studies are grouped into three main use case areas:

1. Genomics and clinical research
2. Diagnostics
3. Treatments and interventions

The connection strengths between the use case areas and the algorithm application categories are illustrated in Figure 1; these were assigned based on the number of proof-of-concept studies associated with each category as well as the applicability of each category to problems typical for a given use case area. It is evident that the category "Processing data with complex structure" is particularly relevant to health and medicine; most of the proof-of-concept studies in this review are based on quantum AI/ML methods.

In the context of quantum AI/ML, variational quantum circuits (VQCs) are sometimes considered to be building blocks of quantum neural networks (QNNs) (Wu et al., 2022), that is, neural networks where parameterized quantum circuits are introduced in the hidden layers. In other instances, a VQC is treated as a synonym for a QNN (along with a parameterized quantum circuit and quantum circuit learning) (Schuld et al., 2021). In this review, no hard distinction is made.

An overview of the explored use cases is given in Figure 2 and a list of the studies and their approaches is provided in Table 1. For many of the proof-of-concept use cases outlined, the quantum approaches are already competitive with the classical benchmarks;

**Figure 2.** Health and medicine quantum computing use cases that have been investigated in proof-of-concept studies.

while many studies have considered downsized versions of the problems, there is generally no reason to suppose that these benefits will not carry over to more realistic problem variants. Moreover, although an entire "quantum algorithm zoo" exists (Quantum Algorithm Zoo, 2022), the algorithms are all based on a limited number of core primitives. Therefore, notwithstanding the particular characteristics of a given problem, such as the data structure, success of applying one algorithm/primitive in a certain field likely bodes well for uses of that algorithm/primitive in other fields. In the following, each study will now be discussed.

### Genomics and clinical research

How can we truly understand an individual at the most granular level? Clearly, genomics is crucial. Over the past decades, we have seen milestones such as the sequencing of the human genome as well as genome-wide association studies (GWAS). It has now become clear, however, that the function and workings of the human genome are much more complex than imagined. The correlations between genomes and outcomes are convoluted and there are, for instance, generally not one-to-one links between genes and diseases. Moreover, pattern problems in the study of haplotypes (groups of genes that are inherited together) and single nucleotide polymorphisms (genomic variations at single base positions between people) quickly become very complicated, reaching nondeterministic polynomial-time (NP) hardness (Lippert et al., 2002).

As a result, there is great interest to adapt the quantum techniques that have already been developed for problems such as string search and matching, for instance, based on Grover's algorithm (Niroula and Yunseong, 2021), to genomic problems. Many experiments have focused on better understanding genomic patterns, leveraging algorithms from the "Processing data with complex structure" and "Search and optimization" categories. For example, DNA sequence alignment was explored with Grover's algorithm (Sarkar et al., 2021) and the quantum Fourier transform (QFT) was applied to pairwise sequence alignment (Prousalis and Konofaos, 2019; Clapis, 2021). De novo DNA sequence reconstruction was carried out through a framework involving the quantum approximate optimization algorithm (QAOA) (Sarkar et al., 2021). Once (genomic) sequences have been obtained, it is then of great interest to analyze the algorithmic information in them; this was explored using Grover's algorithm and phase estimation (Sarkar et al., 2021). Note that many of these early advances in better understanding genomic strings and

Table 1. Overview of quantum algorithms applied in clinical and medical proof-of-concept studies grouped by algorithm application category

| Quantum algorithms | Study focus | References |
| --- | --- | --- |
| **Simulating nature** | | |
| Quantum phase estimation (QPE), variational quantum eigensolver (VQE) | Active space selection and embedding for $F_2$, [Fe] hydrogenase, and temoporfin | Izsák et al. (2022) |
| **Processing data with complex structure** | | |
| Quantum Fourier transform (QFT) | Pairwise sequence alignment | Prousalis and Konofaos (2019), Clapis (2021) |
| Quantum neural network (QNN) | Force fields | Kiss et al. (2022) |
| Quantum generative adversarial network (QGAN) | Generation of molecular entities as drug candidates | Li et al. (2021) |
| Quantum support vector classifier (QSVC) | Virtual screening in drug discovery | Mensa et al. (2023) |
| QSVC, QNN | Classification of molecular descriptor data | Batra et al. (2021) |
| Quantum evolution kernel | Toxicity screening | Albrecht et al. (2023) |
| Quantum determinantal sampling | Clinical data imputation | Kazdaghli et al. (2023) |
| Orthogonal QNN | Classification of retinal color fundus and chest X-ray images | Landman et al. (2022) |
| QFT | Image reconstruction | Kiani et al. (2020) |
| QNN, QSCV | Classification of ischemic heart disease | Maheshwari et al. (2023) |
| Transfer learning-based QNN | Classification of breast cancer | Azevedo et al. (2022) |
| QSVC trained via quantum kernel alignment | Classification of rheumatoid arthritis with thermal hand images | Ahalya et al. (2023) |
| QNN, Quantum distance classifier (QDC) | Classification of Alzheimer's disease | Kathuria et al. (2020), Shahwar et al. (2022) |
| QNN, VQC | Classification of COVID-19 | Sengupta and Srivastava (2021), Yu (2021), Amin et al. (2022), Houssein et al. (2022) |
| QNN | Classification of standardized biomedical images | Cherrat et al. (2022) |
| VQC | Classification of diabetes | Gupta et al. (2022) |
| Quantum random forests, quantum k-nearest neighbors, quantum decision trees, quantum Gaussian Naïve Bayes | Classification of heart failure | Kumar et al. (2021) |
| QDC, QSVC | Classification of bone marrow transplant survival, breast cancer, heart failure | Moradi et al. (2022) |
| VQC, QNN | Classification of states of mind with electroencephalogram (EEG) signals | Aishwarya et al. (2020) |
| Quantum k-means | Classification of heart disease | Kavitha and Kaulgud (2022) |
| QSVC | Classification of medication persistence for individuals with rheumatoid arthritis | Krunic et al. (2022) |
| QNN | Drug response prediction | Sagingalieva et al. (2023) |
| QNN | Treatment effectiveness of knee arthroplasty | Heidari et al. (2022) |
| QNN | COVID-19 outbreak prediction | Kairon and Bhattacharyya (2021) |
| Quantum deep reinforcement learning | Adaptive radiotherapy | Niraula et al. (2021) |
| **Search and optimization** | | |
| Grover's | DNA sequence alignment | Sarkar et al. (2021a) |
| QAOA (quantum approximate optimization algorithm) | De novo DNA sequence reconstruction | Sarkar et al. (2021b) |
| Grover's, QPE | Estimation of algorithmic information from DNA sequences | Sarkar et al. (2021c) |
| QAOA, VQE, VQC, Grover's | Protein structure for lattice model-based systems | Fingerhuth and Babej (2018), Robert et al. (2021), Chandarana et al. (2022), Khatami (2023) |

Table 1. (Continued)

| Quantum algorithms | Study focus | References |
| --- | --- | --- |
| Quantum walk and quantum Markov chain Monte Carlo | Protein structure for nonlattice model-based systems | Allcock et al. (2022), Casares et al. (2022) |
| VQE | Protein–ligand interactions involving lysine-specific demethylase 5 (KDM5A) | Malone et al. (2022) |
| VQE | Binding energy differences for $\beta$-secretase (BACE1) inhibitors | Kirsopp et al. (2022) |
| Gaussian boson sampling* | Ligand binding to the tumor necrosis factor-$\alpha$ converting enzyme | Banchi et al. (2020) |
| VQE | Force fields | Mishra and Shabani (2019) |
| QPE | Electronic structure of cytochrome P450 (CYP) enzyme active sites | Goings et al. (2022) |

*Gaussian boson sampling is a nonuniversal quantum computational method.

sequences may of course also be applied to related problems in omics in due course, for instance, involving RNA sequences. Likewise, the pattern and information encoding perspective can also be complemented with deeper insights at the molecular level through the application of quantum algorithms from the "Simulating nature" and "Search and optimization" categories.

In addition to genomics, there are diverse fields of clinical research in biology and biochemistry that seem poised to benefit from future quantum advantages. The discovery, and ultimately development, of new molecular entities and drugs is of central importance here. While millions of compounds have already been considered in the literature, the total number of possible carbon-based compounds whose molecular masses are similar to those of living systems is around 10^60. Given the large fraction of this gargantuan chemical space that has not yet been explored, the significant potential for future breakthroughs is clear (Dobson, 2004). Multiple overviews about quantum opportunities in the drug discovery space have been published (Cao et al., 2018; Li et al., 2021; Blunt et al., 2022; Santagati et al., 2023). "Simulating nature" algorithm applications play a prominent role, but the other two categories have also been investigated in this context. A general theme is to reduce the need for lengthy and expensive experiments through better simulations of biology, thus creating *in silico* laboratories. The biological molecules and systems that can be modeled with quantum computing today are still relatively small, but these are expected to continually scale as quantum hardware and software further mature.

Protein folding and design have gained much attention in recent years through both classical (Callaway, 2022; Lin et al., 2022) and quantum advances. For instance, lattice model-based systems were explored using variational quantum algorithms, including QAOA (Fingerhuth and Babej, 2018), VQE (Robert et al., 2021), and other variational techniques (Chandarana et al., 2022), as well as Grover's algorithm (Khatami, 2023). For the VQE adaptation (Robert et al., 2021), it was even shown that the number of physical qubits required scales only as the square of the number of amino acids (but without a convergence guarantee), putting structures with 100+ amino acids within reach as quantum hardware develops over the next years (Gambetta, 2022). Generalization to nonlattice models was investigated through quantum walk and quantum Markov chain Monte Carlo methods (Allcock et al., 2022; Casares et al., 2022). Given the classical advances, it is likely that quantum methods will be particularly advantageous for those problem variations that are most challenging for classical methods. For example, this includes trying to predict the structures of proteins with unnatural amino acids (where classical machine learning struggles due to a lack of training data) or trying to understand conformations and behavior in dynamic settings, such as when proteins interact with water molecules, ligands, and other proteins. For protein–ligand interactions, symmetry-adapted perturbation theory (SAPT) was combined with VQE in benchmarks for systems containing the human cancer-relevant protein lysine-specific demethylase 5 (KDM5A) (Malone et al., 2022), and VQE was also extended through density matrix embedding theory (DMET) in order to calculate the binding energy differences for $\beta$-secretase (BACE1) inhibitors (Kirsopp et al., 2022). In addition, docking was investigated through Gaussian boson sampling (Banchi et al., 2020) (a restricted form of quantum computing (Hamilton et al., 2017)) by predicting ligand binding to the tumor necrosis factor-$\alpha$ converting enzyme, which is connected with immune system diseases and cancer.

A variety of further applications to help accelerate the drug discovery process has been explored. These include estimating force fields, accurate calculations of which are crucial for scaling molecular dynamics techniques, through QNNs (Kiss et al., 2022) and VQE (Mishra and Shabani, 2019). In addition, QPE and VQE methods were studied for the active space, the limited number of orbitals that are of primary interest and treated fully quantum mechanically, of (strongly correlated) chemical systems; $F_2$, [Fe] hydrogenase, and the photosensitizer temoporfin were considered (Izsák et al., 2022). Other studies focused on estimating the quantum (and classical) resources to compute the electronic structure of cytochrome P450 enzymes (CYPs) via QPE (Goings et al., 2022) and applying quantum generative adversarial networks (QGANs) to create new drug candidates (Li et al., 2021). Quantum machine learning, specifically quantum support vector classifiers (QSVCs) that enhance calculation of the kernel, also yielded promising results compared with classical state-of-the-art methods for virtual screening in drug discovery (Mensa et al., 2023). In another investigation, cheminformatic molecular descriptor data sets for COVID-19, as well as whole-cell screening sets for plague and *Mycobacterium tuberculosis*, were compressed and then classified using QSVCs and QNN-like methods (Batra et al., 2021). Finally, absorption, distribution, metabolism, excretion, and toxicity (ADMET) studies may be enhanced, as was demonstrated in a toxicity screening experiment where a quantum graph machine learning algorithm (quantum evolution kernel) was applied to a biochemistry data set with information about 286 molecules and their effects on mice (Albrecht et al., 2023).

### Diagnostics

Only when it is possible to accurately assess an individual's health status and potential future development in fine detail can tailored treatments and interventions be properly assigned. As such, quantum computing may enable the move away from (late) diagnoses focused on single diseases toward a regime where a continually updated health status can be determined for each individual; this will only be possible by building on new insights from the previous use case area, "Genomics and clinical research."

Quantum AI/ML algorithms are particularly relevant for diagnostic applications. Not medical care but health-related behaviors, socioeconomic factors, and environmental aspects are now believed to contribute up to 90% to health outcomes (Hood et al., 2016). Hence, it is imperative to understand the quickly growing and increasingly heterogeneous health-relevant data that are becoming available, particularly real-world data (RWD) such as information from electronic health records (EHRs), claims, disease registries, and fitness trackers (Real-World Evidence, 2022). The many potentially pertinent variables lead to high-dimensional feature spaces and interactions between the variables result in complex interdependencies, correlations, and patterns; quantum AI/ML algorithms can penetrate such data structures in ways that are beyond the means of purely classical methods. Furthermore, "Processing data with complex structure" quantum algorithms can even help with enhancing clinical data. For example, quantum determinantal sampling circuits based on Clifford loaders were used to impute a synthetic data set as well as the Medical Information Mart for Intensive Care (MIMIC-III) data set, which contains diagnostic and procedural information for 7,214 patients (Kazdaghli et al., 2023).

Analyzing and getting actionable insights from medical images is a field that has significantly grown in importance over the last years and decades. As such, a broad array of quantum applications is being explored in this space, including the enhancement of processing steps such as image edge detection, segmentation, and classification (Elaraby et al., 2022). In classifying retinal color fundus and chest X-ray images, orthogonal QNNs were investigated, and quantum circuits were also used to accelerate the training of classical neural networks (Landman et al., 2022). Based on computed tomography (CT) and positron emission tomography (PET) data, QFT-based algorithms were developed for enhancing image reconstruction (Kiani et al., 2020). QNNs and QSCVs were applied to EHRs to classify ischemic heart disease (Maheshwari et al., 2023), while transfer learning-based QNNs were explored in the context of classifying breast cancer (Azevedo et al., 2022). Rheumatoid arthritis was detected by classifying thermal hand images with QSVCs trained via quantum kernel alignment (Ahalya et al., 2023). Alzheimer's disease was classified with MRI images using QNNs (Shahwar et al., 2022), and COVID-19 was classified with QNNs using chest X-ray (Houssein et al., 2022) as well as CT lung images (Sengupta and Srivastava, 2021; Amin et al., 2022). Finally, quantum transformers were explored to achieve more efficient neural network architectures for classifying standardized biomedical images; the quantum architectures only required thousands of parameters, compared with millions for the best classical approaches (Cherrat et al., 2022).

Next to images, diseases and disease risks have also been classified and predicted in early studies of supervised quantum AI/ML. Copy number variations (CNVs), differences in the number of repetitions of a genomic section between individuals, in neuronal single-cell samples from healthy individuals and those with Alzheimer's disease were used as features; building on the efficiency with which quantum computers can evaluate inner products, this allowed quantum distance classifiers (QDCs) to predict whether a given sample is from a healthy or a sick individual (Kathuria et al., 2020). COVID-19 was diagnosed through VQCs based on features such as temperature (fever), fatigue, muscle pain, and coughing (Yu, 2021). VQCs were also employed to predict diabetes (Gupta et al., 2022). A diversity of methods – quantum random forests, quantum k-nearest neighbors, quantum decision trees, and quantum Gaussian Naïve Bayes – was studied for the purpose of classifying heart failure (Kumar et al., 2021). Conversely, QDCs and QSVCs were applied to assess multiple conditions in the same study, namely bone marrow transplant survival, breast cancer, and heart failure (Moradi et al., 2022). Moreover, VQCs as well as QNNs were even used to predict states of mind based on electroencephalogram (EEG) signals from individuals who responded toward a product with a like/dislike in a neuromarketing experiment (Aishwarya et al., 2020). Finally, in the same way that unsupervised learning is a younger discipline than supervised learning for classical ML, the field of unsupervised quantum AI/ML is younger than supervised quantum AI/ML. Still, even here there is already early medical work underway – the quantum k-means algorithm was used for clustering individuals based on their demographic and laboratory measurement data and predicting heart disease (Kavitha and Narasimha, 2022).

### Treatments and interventions

The applications outlined in the previous two use case areas, "Genomics and clinical research" as well as "Diagnostics," form the foundation for tailored treatments and interventions. As for diagnostics, quantum AI/ML algorithms lend themselves particularly well to treatment and intervention use cases.

Next to knowing an individual's health status and disease risks, it is essential to understand likely adherence, engagement, and behavior in order to achieve optimal outcomes (Dentzer, 2013). RWD again plays a central role here. Based on EHRs, for instance, the medication persistence of individuals with rheumatoid arthritis was predicted with QSVCs and a general framework to help assess empirical quantum advantage potential was introduced (Krunic et al., 2022). Another essential research topic on the road to precision medicine is treatment effectiveness. In one study, drug response was predicted by deriving $IC_{50}$ values (the drug concentrations where the response is half of the maximum) using QNNs (Sagingalieva et al., 2023). In addition, for the purpose of forecasting knee arthroplasty QNNs were applied to clinico-demographic data from 170 individuals that were treated over two years. The results were encouraging, but the study also noted that further validation using unstructured RWD is needed (Heidari et al., 2022). Optimal measures at the population level require better models too, for example, regarding outbreak prediction and disease spread dynamics. Using a COVID-19 time series data set with confirmed cases, number of deaths, and number of recovered individuals, different types of QNNs (including continuous-variable ones) were applied for this purpose (Kairon and Bhattacharyya, 2021).

As quantum techniques continue to mature and proliferate, there is hope that they can accelerate the discovery process itself as well as enable progress for some of the thorniest medical treatment and intervention problems. Precision oncology is a case in point. Currently, only a third of individuals respond to drug-based cancer therapies (Spilker, 2022). One key challenge is the need to make sense of terabytes and terabytes of relevant data for an individual with cancer. Work has already begun on leveraging quantum algorithms for the purpose of getting actionable insights from such data and ultimately tailoring cancer treatments to the level of the individual (Abbott, 2021). One of the early applications showing promise is adaptive radiotherapy, as was demonstrated by modeling the clinical decisions as quantum states and applying quantum deep reinforcement learning to an institutional data set based on 67 stage III nonsmall cell lung cancer patients (Niraula et al., 2021). Yet another research frontier concerns the intersection of quantum algorithms with single-cell technologies with the aim to enhance the development of cell-centric therapeutics (Basu et al., 2023).

Table 2. General achievements of quantum computing in healthcare, medicine, and life sciences as well as other fields

| Success | Description |
| --- | --- |
| Competitive results | As evidenced by the literature covered in this review, quantum algorithms are on the cusp of providing benefits (in terms of accuracy, calculation speed, energy efficiency, or input data requirements) for certain medical applications |
| Development of innovation ecosystems | The commercialization of quantum computing has led to the creation of a multiplicity of quantum hubs and valleys across the globe, bringing together large and small organizations to drive cutting-edge research and shaping the future of medicine (Murphy and Douglas, 2023, QuantumBasel brings the first commercially viable physical quantum computer to Switzerland, 2023) |
| Improvements based on quantum-related computing methods | In conjunction with (universal) quantum computing research, quantum annealing (Quantum in Life Sciences: The Future is Now, 2023) and quantum-inspired (classical) (Buntz, 2023) methods are being enhanced, enabling additional computational benefits in healthcare and life sciences |
| Progress of other quantum technologies | The principles of quantum mechanics are not just useful for computational purposes but are also employed to achieve quantum-safe communication and quantum sensors with unprecedented sensitivities, where securing medical data and measuring biological signals are key applications respectively (Flöther and Griffin, 2023) |
| Public fascination | Quantum computing, with its counterintuitive foundations, has sparked public interest with many people looking to learn more about and enter the field, schools and universities offering new courses, and medical fields being reimagined (Flöther, forthcoming) |

## Conclusion and perspective

Ever since the beginnings of medicine thousands of years ago, medicine has continually incorporated new ideas, knowledge, and methods to become more effective. Quantum computing is very young but, as the only known computational model that has exponential speedups compared with traditional approaches (National Academies of Sciences, Engineering, and Medicine, 2019), poised to become a mighty tool in healthcare and medicine with the power to make previously intractable problems now solvable. Despite its youth, quantum computing has already achieved a number of general successes, as summarized in Table 2.

For quantum computing to become this powerful enabler for health and medicine and for a wide range of quantum-enhanced solutions to go into production, however, a wide range of technical and ethical challenges must still be overcome (Figure 3). First, quantum hardware and software need to continue improving, including more efficient algorithms, decreased error rates, and increased qubit numbers. Second, there are various challenges around making quantum computing practical for medicine which are similar to those in digital health efforts. These include data accessibility (without which even quantum computing cannot wield its power), model explainability (essential for obtaining the support of clinicians, medical practitioners, and individuals), and patient privacy (critical for developing the long-term trust of individuals in the technology). Third, new challenges specific to quantum computing have appeared. Examples are data security, replicability, and skill development, which will now be discussed in turn.

Some quantum algorithms, specifically Shor's and Grover's algorithm, are able to solve the mathematically hard problems at the heart of current cryptography significantly faster than classical methods. All data that are not encrypted with quantum-safe protocols are thus already at risk due to the possibility of "harvest now, decrypt later" attacks (Harishankar et al., 2023); given the sensitivity and long security time value of medical data, this problem is exacerbated. As a result, cross-industry quantum-safe standards are already being developed (NIST Announces First Four Quantum-Resistant Cryptographic Algorithms, 2022) and will soon be implemented (Migrating to Post-Quantum Cryptography, 2022). Furthermore, replicability, required in order to achieve clinical approvals and individual acceptance, is a challenge for quantum computers. Quantum computers, by their very nature, are designed to go beyond traditional means and address classically intractable problems; for many problems, however, new (quantum) solutions cannot be efficiently verified. Replicability is further complicated by the probabilistic nature of quantum computing, the multifarious architectures, the presence of noise, and the (still) limited access to quantum hardware. Hence, methodologies and frameworks to secure regulatory approvals and general support will need to be developed, as has been done for classical AI/ML (Benjamens et al., 2020). Finally, there is fierce competition for quantum talent, particularly practitioners who combine quantum skills with medical expertise. As a result, talent development needs to be extended, including the introduction of new roles such as "quantum translators" (Mohr et al., 2022).

The development of medicine-focused quantum computing collaborations and consortia is critical with regard to addressing many of these challenges. Such ecosystems are beginning to emerge (Zinner et al., 2021a, 2021b; Major investment for developing Denmark's first fully functional quantum computer, 2022; Cleveland Clinic and IBM Begin Installation of IBM Quantum

| | | | |
|---|---|---|---|
| **General quantum computing challenges** | **Algorithm efficiency** Develop more efficient quantum algorithms that can handle difficult data sets (high dimensionality, noise, small or large size, …) | **Errors** Reduce calculation errors through better physical devices and software (error suppression, mitigation, and correction) | **Number of qubits** Increase number of qubits, e.g. to enable encoding of more features |
| **General digital health challenges** | **Data accessibility** Ensure easy access to high-quality clinical and real-world data sets | **Model explainability** Avoid black-box approaches and provide intelligible reasoning as to why a model came up with a given answer in a given setting | **Patient privacy** Allow each individual to have full control over access to their medical data |
| **Medical quantum computing challenges** | **Data security** Secure medical data long-term by employing quantum-safe cryptography | **Replicability** Achieve sufficient degree of replicability (despite the limits of quantum simulators, multifarious architectures, noise, …) | **Skill development** Enable practitioners to combine medical with quantum expertise, including new roles such as "quantum translators" |

**Figure 3.** Examples of technical and ethical challenges that must be addressed for quantum computing to become transformative in health and medicine.

System One, 2022; uptownBasel opens first quantum computer hub for commercial use in Switzerland, 2022) to help practitioners tackle problems with a quantum state of mind. In healthcare, there has been much discussion about the journey towards precision medicine and the quadruple aim (better health, lower costs, enhanced patient experiences, and improved healthcare practitioner work lives) (Bodenheimer and Christine, 2014). While a range of technical and ethical challenges remain, quantum computing is poised to become a key enabler for advancing towards the holy grail: keeping people healthy through proactive medical care and guidance at the level of an individual. All of this will take time and effort, but the significant rewards along the road toward quantum-enhanced health and medicine make it a highly worthwhile journey to start sooner rather than later.

## Methods

The literature search was conducted primarily through the Google Scholar and PubMed platforms. The logical operators OR and AND were combined with search terms such as the following: AI, algorithm, application, artificial intelligence, biology, chemistry, clinical, clinical research, diagnosis, diagnostics, drug, genomics, health, intervention, machine learning, medicine, ML, nature, optimization, QML, quantum, quantum AI, quantum artificial intelligence, quantum computing, quantum machine learning, search, simulation, and treatment.

Studies were only included if the work explored quantum computing algorithms for applications within, or closely related to, health and medicine. Moreover, the focus of this review is on the quantum circuit model and gate-based quantum computers. Gaussian boson sampling is briefly touched on, but other nonuniversal approaches, such as quantum annealing, were excluded.

**Acknowledgments.** The author would like to thank Travis L. Scholten for helpful discussions.

**Competing interests.** The author declares no conflicts of interest.

## Connections references